\documentclass[aps,prb,showpacs,floatfix]{revtex4}
\usepackage{graphicx,amssymb}

\begin{document}

\title{Theory of the microwave induced zero resistance states in two-dimensional electron systems}

\author{S. A. Mikhailov}
\email[E-mail: ]{sergey.mikhailov@physik.uni-augsburg.de}

\affiliation{Institute of Physics, University of Augsburg, D-86135 Augsburg, Germany}

\date{\today}

\begin{abstract}
The phenomena of the microwave induced zero resistance states (MIZRS) and the microwave induced resistance oscillations (MIRO) were discovered in the ultraclean two-dimensional electron systems in 2001 -- 2003 and have attracted great interest of researchers. In spite of numerous theoretical efforts the true origin of these effects remains unknown so far. We show that the MIRO/ZRS phenomena are naturally explained by the influence of the ponderomotive forces which arise in the near-contact regions of the two-dimensional electron gas under the action of microwaves. The proposed analytical theory is in agreement with all experimental facts accumulated so far and provides a simple and self-evident explanation of the microwave frequency,  polarization, magnetic field, mobility, power and temperature dependencies of the observed effects.
\end{abstract}

\pacs{73.43.Qt, 73.50.Mx, 52.35.Mw, 73.40.Cg  }





 
\maketitle

\section{Introduction\label{intro}}

\begin{figure}
\includegraphics[width=12cm]{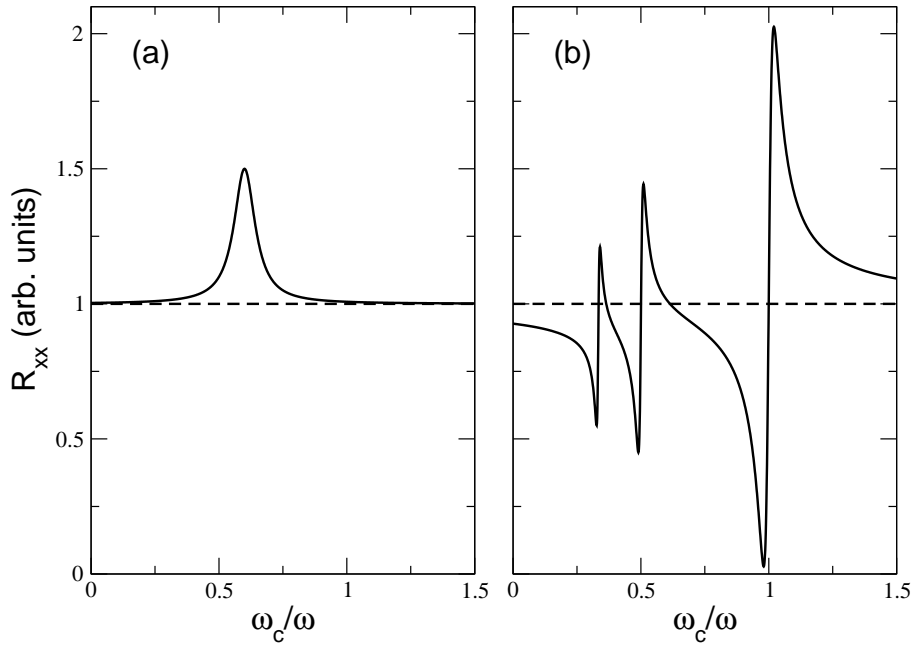}
\caption{\label{compar}This picture schematically shows the main features and the main differences of the microwave induced photoresistance response in (a) the moderate mobility samples \cite{Vasiliadou93}  and (b) the very-high-mobility samples \cite{Mani02,Zudov03,Dorozhkin03,Yang03,Kovalev03,Willett04,Zudov04,Mani04,Mani04b,Mani04c,Studenikin04,Du04,Smet05,Studenikin05,Mani05,Yang06,Zudov06a,Zudov06b,Studenikin07,Andreev08,Hatke09}. In the moderate mobility samples the resonance at $\omega_c=\omega_c^{res}$ {\em is shifted} from the cyclotron $\omega_c=\omega$ to the magnetoplasmon frequency $\omega_c^{res}=(\omega^2-\omega_p^2)^{1/2}$ and the photoresistance is an {\em even} function of $\omega_c-\omega_c^{res}$. In the very-high-mobility samples one observes multiple cyclotron resonances at $\omega_{c}=\omega_{c,k}^{res}=\omega/k$ and the photoresponse is an {\em odd} function of $\omega_c-\omega_{c,k}^{res}$. Here $k$ is integer and $\omega_p$ is the plasma frequency (\ref{plfreq}). The dashed lines schematically show the dark magnetoresistance (without microwaves). }
\end{figure}

The discovery of the microwave induced zero resistance states (MIZRS)  \cite{Mani02,Zudov03} in two-dimensional electron-gas (2DEG) systems has aroused great interest of the physical community \cite{PhysToday03}. The MIZRS effect is observed in standard Hall-bar GaAs-AlGaAs quantum-well structures placed in the perpendicular magnetic field $B$. In the absence of microwaves one sees the conventional picture of the quantum Hall effect and the Shubnikov-de Haas oscillations in quantizing magnetic fields and the corresponding classical behavior ($R_{xy}\propto B$, $R_{xx}\approx const$) at low $B$; here $R_{xy}$ and $R_{xx}$ are the Hall and the diagonal resistance respectively. The irradiation of the samples by microwaves (with the frequency $f\simeq 50-100$ GHz) leads to {\em giant oscillations} of the diagonal magnetoresistance $R_{xx}$ at low (classical) magnetic fields $B\lesssim 0.5$ T. The effect is governed by the ratio $\omega/\omega_c$ of the microwave $\omega=2\pi f$ to the cyclotron frequency $\omega_c=eB/m^\star c$ ($m^\star$ is the electron effective mass in GaAs). In the magnetic field intervals  corresponding to the conditions $k-1/2\lesssim \omega/\omega_c<k$, $k=1,2,\dots$, a dramatic growth of the magnetoresistance is observed (by a factor of $6-10$), while at $k< \omega/\omega_c\lesssim k+1/2$ it is strongly suppressed down to ``zero-resistance'' states $R_{xx}\approx 0$, Figure \ref{compar}(b). Exactly at $\omega=k\omega_c$ there is no change of the resistance, $\delta R_{xx}\approx 0$. The largest amplitudes of the $R_{xx}$ oscillations are seen near the fundamental cyclotron harmonics $k=1$, however the oscillation amplitudes fall down rather slowly with $k$ and the oscillations remain quite visible up to $k\simeq 10$.  No apparent microwave induced changes are  observed in the Hall resistance $R_{xy}$. In the Corbino geometry very large {\em conductance} oscillations and zero conductance states have been also seen \cite{Yang03}. 

In the pioneering \cite{Mani02,Zudov03} and the subsequent experimental papers \cite{Dorozhkin03,Yang03,Kovalev03,Willett04,Zudov04,Mani04,Mani04b,Mani04c,Studenikin04,Du04,Smet05,Studenikin05,Mani05,Yang06,Zudov06a,Zudov06b,Studenikin07,Andreev08,Bykov08,Tung09,Hatke09,Konstantinov09,Konstantinov10,Bykov10a,Bykov10b,Wiedmann10}  the MIZRS effect has been observed in GaAs-AlGaAs quantum wells with an {\em extremely high electron mobility} $\mu\gtrsim 2\times 10^7$ cm$^2$/Vs.  A very similar effect but with smaller oscillation amplitudes -- the microwave induced resistance oscillations (MIRO) -- has been found \cite{Zudov01,Ye01} two years earlier in the lower quality samples with the mobility $\mu\simeq 3\times 10^6$ cm$^2$/Vs. In samples with $\mu\lesssim 10^6$ cm$^2$/Vs the MIROs have not been observed (for the only exception see Ref. \cite{Bykov06}). In such samples, instead, the photoresistance measurements demonstrate a {\em completely different} photoresponse \cite{Vasiliadou93}: a weak Lorenzian peak  corresponding to the excitation of the 2D magnetoplasmon, Figure \ref{compar}(a). In the first  MIRO experiments \cite{Zudov01,Ye01} ($\mu\simeq 3\times 10^6$ cm$^2$/Vs) the magnetoplasmon resonance has been seen together with the $\omega_c$-related resistance oscillations.

The discovery of the MIZRS has caused an avalanche of theoretical publications \cite{Dorozhkin03,Durst03,Lei03a,Shi03,Andreev03,Phillips03b,Ryzhii03a,Ryzhii03b,Vavilov04,Volkov03b,Ryzhii03e,Shikin03,Koulakov03,Bergeret03,Dmitriev03,Dmitriev04,Ryzhii04b,Park04,Dmitriev05,Ryzhii05,Inarrea05,Auerbach05,Inarrea06,Volkov07,Chepelianskii07,Dmitriev07a,Dmitriev07b,Inarrea07,Inarrea08,Wang08,Dmitriev09,Finkler09,Inarrea10,Mikhailov03c,Chepelianskii09}. Several different scenarios for the explanation of the effect have been put forward. Among the proposed ideas are the so called ``displacement'' model \cite{Ryzhii03a,Ryzhii03b,Shikin03,Durst03,Lei03a,Shi03,Vavilov04}, originally proposed in Refs. \cite{Ryzhii70,Ryzhii86} for a different physical situation, the microwave induced dynamical symmetry breaking  \cite{Andreev03}, the ``inelastic'' model \cite{Dorozhkin03,Dmitriev03,Dmitriev05,Dmitriev07a,Dmitriev07b}, nonparabolicity effects \cite{Koulakov03}, the photon assisted quantum tunneling \cite{Shi03}, phase transitions caused by electron pairing due to the exciton exchange \cite{Mani02} and a quantum model that involves the prime number theorem \cite{Phillips03b}. The ``displacement'' model suggests, for example, that microwaves assist the impurity and phonon scattering which leads to the spatially indirect transitions between the Landau levels $N$ and $N'=N+k$, $k=1,2,3,\dots$. The ``inelastic'' mechanism supposes that the Landau levels are substantially broadened due to disorder and that the microwave absorption modifies the distribution of electrons over the broadened Landau levels. 

The vast majority of the proposed theories \cite{Dorozhkin03,Durst03,Lei03a,Shi03,Andreev03,Phillips03b,Ryzhii03a,Ryzhii03b,Vavilov04,Volkov03b,Ryzhii03e,Shikin03,Koulakov03,Bergeret03,Dmitriev03,Dmitriev04,Ryzhii04b,Park04,Dmitriev05,Ryzhii05,Inarrea05,Auerbach05,Inarrea06,Volkov07,Chepelianskii07,Dmitriev07a,Dmitriev07b,Inarrea07,Inarrea08,Wang08,Dmitriev09,Finkler09,Inarrea10} has been looking for the MIZRS origin in the {\em bulk} of the 2DEG. They have considered an {\em infinite} 2DEG system and have examined the influence of microwaves on the bulk resistivity $\rho_{xx}$  of the 2D gas. We have already emphasized in Refs. \cite{Mikhailov03c,Mikhailov04a} that ignoring the fact that the real experimental samples have finite dimensions one cannot get an adequate description of the real experiments. Indeed, consider an infinite 2DEG under the action of the external microwave field $E_x^0e^{-i\omega t}$. In the infinite sample the only relevant parameter which may be compared to $\omega$ is the cyclotron frequency $\omega_c$. Therefore the system response is determined by the ratio $\omega/\omega_c$, for example the induced electric current
\begin{equation}  
j_x(t)=\sigma_{xx}(\omega)E_x^0\propto  \frac{E_x^0}{\omega^2-\omega_c^2}\label{curr1}
\end{equation} 
has a resonance at the cyclotron frequency. Due to the same reason different bulk scenarios give the microwave induced resistivity changes determined by functions like $ -\sin(2\pi\omega/\omega_c)$ (e.g. in Refs. \cite{Durst03,Vavilov04}) or 
$$\sum_kJ_k\left(const\frac{E_x^0}{\omega_c^2-\omega^2}\right)$$
(e.g. in Refs. \cite{Park04,Inarrea05,Inarrea06,Inarrea07,Inarrea08}; here $J_k$ are the Bessel functions). 

In the {\em finite-size} 2DEG samples (e.g. in the Hall bars of the width $W$ in the $x$-direction) a new frequency parameter appears -- the plasma frequency 
\begin{equation} 
\omega_p\approx \left(\frac{2\pi^2n_se^2}{m^\star  \kappa W}\right)^{1/2},
\label{plfreq} 
\end{equation} 
where $n_s$ is the 2D electron density and $\kappa$ is the dielectric constant of the surrounding medium. In such samples the internal microwave electric field $E_x$ {\em really acting} on the electrons differs from the external field $E_x^0$ due to the screening, 
\begin{equation} 
E_x=\frac{E_x^0}{\zeta(\omega)},\ \ {\rm where} \ \ \zeta(\omega)\approx 1-\frac{\omega_p^2}{\omega^2-\omega_c^2}
\label{zeta}
\end{equation} 
is the effective dielectric function of the 2DEG stripe (see, e.g. Ref. \cite{Mikhailov04a}). 
The response current (\ref{curr1}) then assumes the form
\begin{equation}  
j_x(t)=\sigma_{xx}(\omega)E_x=\frac{\sigma_{xx}(\omega)}{\zeta(\omega)}E_x^0\propto  \frac{E_x^0}{\omega^2-\omega_c^2-\omega_p^2}
\label{curr2}
\end{equation} 
so that the cyclotron resonance is shifted to the magnetoplasmon frequency 
\begin{equation} 
\omega_{mp}=(\omega_p^2+\omega_c^2)^{1/2}
\label{magnetopl}
\end{equation} 
(the so called depolarization shift). Similarly, in all formulas for $\rho_{xx}$  suggested in the bulk-scenario papers the input microwave field $E_x^0$ should be replaced by the screened field (\ref{zeta}), $E_x^0\to E_x^0/\zeta(\omega)$. This would then shift all the $\omega_c$-related resistivity oscillations and would completely destroy any seeming agreement with experiments (the depolarization shift is not negligible under the real experimental conditions, see a detailed comparison in Table I of Ref. \cite{Mikhailov04a}; to the  contrary, it is very large and has been many times observed both in the absorption \cite{Demel88,Demel91,Kukushkin06} 
experiments). 

Moreover, one should not ignore that (already mentioned) fact that about ten years before the first MIZRS experiment by Mani et al. \cite{Mani02} a very similar microwave photoresistance experiment was performed in the same experimental group by Vasiliadou et al. \cite{Vasiliadou93}. The only difference between the old (1993) and new (2002) experiments was the mobility of the samples ($\mu\lesssim 10^6$ cm$^2$/Vs in 1993 and $\mu\simeq 2\times 10^7$ cm$^2$/Vs in 2002); all other parameters, such as the electron density, the sample dimensions, the microwave frequency, the temperature, were the same. But the results of these two experiments turned out to be amazingly different. Vasiliadou et al. observed a weak magnetoplasmon resonance (\ref{magnetopl}), Figure \ref{compar}(a), which results from the screening of the external field (\ref{zeta}) and is well understood. It was natural to expect that in the higher-mobility samples the magnetoplasmon resonance will be only slightly modified, e.g. its linewidth will be reduced. In contrast, the giant $\omega_c$-related oscillations have been discovered, Figure \ref{compar}(b). A correct and comprehensive theory should evidently answer the questions: {\em Why the well-known screening effect does not manifest itself in the MIZRS experiments? Why the standard magnetoplasmon picture is easily observed in samples with relatively low mobility, while the very-high-mobility samples demonstrate a completely different response? Why the giant $R_{xx}$ oscillations and zero resistance states are not seen in low-mobility samples?} These questions cannot be answered within the bulk models, since the depolarization shift is a consequence of the finite width of the sample. The main puzzle of the MIZRS phenomenon have not thus been even considered in the bulk-scenario papers. 

The bulk theories of MIZRS/MIRO have been also shown to be in disagreement with experimental facts obtained in the very important paper of Smet et al. \cite{Smet05}. Being strongly related to the optical transitions between the Landau levels, the ``bulk'' scenarios are intrinsically sensitive to the sense of the circular polarization of the incident microwave radiation. If the ``bulk'' approaches were valid, the MIZRS effect, like the absorption, had to be seen at the active circular polarization of radiation and completely disappear at the inactive polarization. In the experiment of Smet et al. \cite{Smet05} it has been found that the amplitudes of the microwave induced resistance oscillations at the right- and left-circularly polarized waves are essentially the same, while the absorption spectra showed a very large difference. 

The puzzle of the MIZRS/MIRO phenomena has thus been remaining unexplained so far.

Before proceeding to the presentation of our theory let us enumerate a few  further specific features of the discussed phenomena.

First, the MIZRS/MIRO effects are observed under the conditions 
\begin{equation} 
\hbar/\tau\ll T\simeq \hbar\omega_c\lesssim\hbar\omega\ll E_F,\ \ \ N\simeq 50-100,
\label{conditions}
\end{equation} 
which suggest that they must have a {\em classical} origin; here $\tau$ is the momentum relaxation time extracted from the mobility, $T$ is the temperature, $E_F$ is the Fermi energy and $N$ is the number of occupied Landau levels. 
Second, the MIRO are seen around $\omega\simeq k\omega_c$ with the harmonics number $k$ up to $k\simeq 10$. It is known 
that the inter-Landau-level transitions $N\to N'$ are usually strictly forbidden except for the case $N\leftrightarrow N\pm 1$. This selection rule could be violated only if the ac electric field acting on the electrons was {\em strongly inhomogeneous} on the scale of the cyclotron radius $R_c$. The value of $R_c$ under the actual experimental conditions varies from one to a few microns which is about two orders of magnitude smaller than both the wavelength of radiation ($\gtrsim 1$ mm) and the sample dimensions ($\gtrsim 0.5$ mm). This completely excludes the chance to bypass the selection rule $N'-N=\pm 1$. The idea that this rule could be broken by impurities and disorder can hardly save the situation in the ultraclean samples where disorder (if essential at all) should be smooth. 

In addition, it is known that in the strongly inhomogeneous external fields one should expect the Bernstein modes \cite{Gudmundsson95} at the frequencies corresponding to the intersection of the magnetoplasmon (\ref{magnetopl}) with the cyclotron harmonics $\omega=k\omega_c$. Usually, the Bernstein modes are very weak and seen, if at all, at $k\simeq 2,3$. In the MIRO-MIZRS and other similar absorption/magnetotransport experiments no indications on the Bernstein modes have been observed.

Among the further puzzles of the MIRO/ZRS phenomena one should mention the observation, under certain conditions, of the apparent {\em negative} resistance in the finite intervals of $B$, Ref. \cite{Willett04}, and the suppression of the MIRO oscillations by the parallel magnetic field $B_\parallel$, Ref. \cite{Yang06}. The latter effect is observed in the quite moderate fields $B_\parallel\sim 1$ T, when the spin effects are still not to be expected.

Summarizing the state of the art in the MIRO/ZRS research one has to admit the presence of a very large number of  clear experimental data, which appear to be, however, in contradiction with common sense, and the absence of a theory which could reasonably explain them. On the other hand, the absolute clarity and the large amplitude of the observed oscillations suggest that the true explanation should actually be very simple and self-evident. Such explanation is given in the present paper.

\section{Explanation of the MIRO/ZRS experiments\label{sec-explain}}

\begin{figure}
\includegraphics[width=12cm]{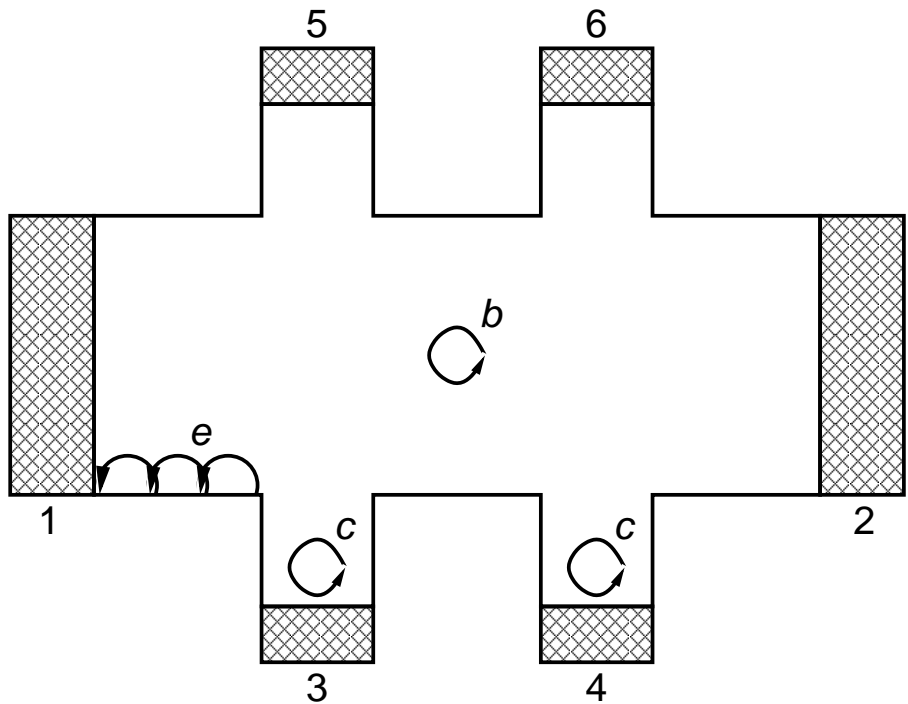}
\caption{\label{hallbar}A schematic view of the Hall-bar sample. The circles and semicircles show electrons rotating around the cyclotron orbits in the bulk ($b$) and in the near-contact ($c$) regions, as well as the skipping electron orbits near the edge ($e$) of the sample.
}
\end{figure}

\paragraph*{1.} 
There exist three types of electrons in a Hall-bar sample, Figure \ref{hallbar}: in the bulk (`$b$'), near the edge (`$e$')  and in the near-contact regions (`$c$'). In the {\em  absorption} experiments the main contribution to the measured signal is given by the bulk electrons `$b$'.  They move in the screened electric field (\ref{zeta}) and absorb the microwave energy at the magnetoplasmon frequency (\ref{magnetopl}). Since the field in the bulk is weakly inhomogeneous on the cyclotron radius scale, the absorption spectra do not demonstrate any nonlocal resonances (Bernstein modes). 

In the {\em transport} experiments made on the samples with a moderate electron  mobility $\mu\lesssim 10^6$ cm$^2$/Vs the magnetoresistance response also shows the resonance at the magnetoplasmon frequency (\ref{magnetopl}), Figure \ref{compar}(a). This evidently points to the bulk mechanism of the  photoresponse in this case: electrons resonantly absorb radiation at $\omega=\omega_{mp}$, the temperature of the electron gas increases and the bulk resistivity of the 2DEG changes. 

The dramatic change of the photoresponse in the ultraclean samples suggests that, {\em in addition} to the conventional bulk photoresistance mechanism, a new mechanism comes into play in such samples. This is supported by the fact that in the MIRO regime one can sometimes observe {\em both} the magnetoplasmon resonance {\em and} the $\omega_c$-related oscillations. Therefore we do  not attempt to find the reason of the new effect in the scattering mechanisms of the `$b$' electrons, but will search for another, bulk-unrelated contribution to the measured magnetoresistance of the sample.

\paragraph*{2.}  The origin of the second contribution to the photoresistance signal may lie  near the edge or in the near-contact regions of the sample. The edge mechanisms of MIZRS and MIRO have been discussed in Refs. \cite{Mikhailov03c,Chepelianskii09}. They cannot however explain all experimentally observed features since the edge electrons `$e$', Figure \ref{hallbar}, are practically under the same conditions as the `{\em b}' electrons (for example the field near the edge is weakly inhomogeneous on the cyclotron-radius scale, therefore it would hardly be possible to explain the higher cyclotron harmonics in the MIZRS effect). Consider the near-contact electrons `$c$'. They also feel not the external but the screened microwave electric field. In the very vicinity of the contacts, however, the field is screened not by the electrons of the 2DEG but by those of the metallic contact. Since the electron density in the metal is many orders of magnitude larger than in the 2DEG, the screening by the contacts in the near-contact regions is much more efficient. 

\begin{figure}
\includegraphics[width=12cm]{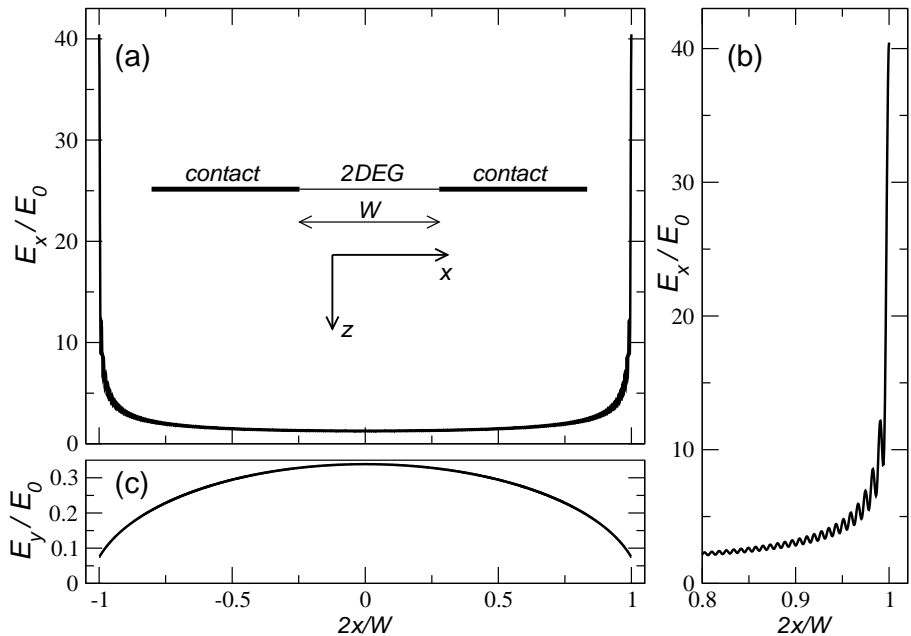}
\caption{\label{screen} The electric field inside the gap between the contact wings, in the absence of the 2D electrons, at $\eta_c\equiv 2\pi\sigma_c/c\sqrt{\kappa}=100$ and $W/\lambda=0.1$ (in a typical experiment the wavelength of radiation $\lambda\simeq 3$ mm and $W\simeq 0.3$ mm). 256 Fourier harmonics have been used. (a) and (b) -- electric field is perpendicular to the boundary 2DEG -- contact, (c) -- electric field is parallel to the boundary 2DEG -- contact. Inset in (a) shows the geometry of the 2D stripe between the two contact wings.}
\end{figure}

The influence of the contacts on the microwave response of a 2DEG stripe has been studied in Ref. \cite{Mikhailov06a}. In that paper the distribution of the electric field in the system ``contact -- 2DEG -- contact'' (see the inset to Figure \ref{screen}(a)) is found from the solution of the Maxwell equations by expanding the fields in Fourier series over the functions $\sim \cos(2\pi x n/W)$. The contacts are described as infinitely thin 2D layers with a large (real) surface conductivity $\sigma_c$. Figure \ref{screen} shows the thus calculated electric fields $E_x(x,z=0)$ and $E_y(x,z=0)$ in the gap $|x|\le W/2$ between the contacts in the absence of the 2D electrons (i.e. only the screening by the metallic contacts is taken into account). A very large number $N_f$ of Fourier harmonics ($N_f=256$ in Figure \ref{screen}) is required to get a reasonable convergency of the results for the field $E_x$. Still remaining small oscillations in Figure \ref{screen}(a,b) disappear when $N_f$ is further increased. The calculations of the field $E_y$ parallel to the boundary 2DEG -- contact converge much faster with $N_f$. For further details of calculations see Ref.  \cite{Mikhailov06a}.

Three important features are seen in Figure \ref{screen}. 

First, the amplitude of the electric  field $E_x(x)$ polarized perpendicularly to the boundary 2DEG -- contact is {\em much larger} than that of the incident wave $E_0$. The field of the incident electromagnetic wave {\em is strengthened} by the metallic contact in the near-contact region. This is a well known effect resulting from  the induced charge accumulation near sharp edges of metallic objects, Ref. \cite{Landau8}, \S 3. If the metallic layer is infinitely thin and the conductivity of the metal is infinitely large (an ideal metal) the field $E_x(x)$ diverges as $1/\sqrt{(W/2)^2-x^2}$ near the edges, Ref. \cite{Landau8}, \S 3. In the real system with a finite $\sigma_c$ and a finite thickness of the metallic contact in $z$ direction  the divergency is cut off but the near-contact field $E_c\equiv E_x(x=\pm W/2)$ will still be much larger than the field of the incident wave, $E_c\gg E_0$. 

Second, the near-contact electric field $E_x(x)$ is {\em strongly inhomogeneous} on the cyclotron radius scale. In the experiments $R_c$ varies between $1$ and $10$ $\mu$m, while the only length $l$ relevant to the cut-off of the field divergency is of order of $0.1$ $\mu$m (the thickness of the AlGaAs layer on top of the 2DEG). This explains the fact that the very large number of the cyclotron harmonics is seen in the MIRO/ZRS experiments. Notice that the field $E_x(x)$ in the bulk, as well the field $E_y(x)$ in the whole sample are quasi-uniform on the scale of $R_c$. 

Third, the near-contact electric field is {\em strongly linearly polarized} near the contact. The field $E_y$, parallel to the boundary contact -- 2DEG, is {\em smaller} than the incident field amplitude $E_0$. It is screened by the self-inductance effect caused by the tangential current $j_y$ induced in the contact by the incident radiation. In contrast, the $E_x$ component of the field is {\em much larger} than $E_0$ in the near-contact region. That is, independent of the sense of the circular polarization of the incident wave the near-contact electrons  feel the {\em linearly polarized} ac electric field. This explains the insensitivity of the MIZRS effect to the sense of the circular polarization of microwaves observed by Smet et al. \cite{Smet05}. 

\paragraph*{3.} Thus, near the contacts  the 2D electrons move in the {\em strong, strongly inhomogeneous and linearly polarized} ac electric field. What physical effect can then lead to the observed photoresistance features? Remind that the MIZRS effect is seen in the very-high-mobility, {\em ultraclean} samples. Therefore we assume that the electron scattering {\em is not important in the discussed phenomena at all} and that a {\em collisionless effect} should be responsible for the MIRO/ZRS phenomenon. This idea is supported by comparing Figures \ref{compar}(a) and \ref{compar}(b). The magnetoplasmon absorption resonance in Figure \ref{compar}(a) is described by an {\em even} function with respect to the resonance frequency $\omega=\omega_{mp}$, i.e. it is determined by the {\em real} part of the effective dynamic conductivity of the system $\sigma'_{xx}(\omega)$. In contrast, the MIRO/ZRS resonances are evidently described by {\em odd} functions of $\omega-k\omega_c$, Figure \ref{compar}(b), which strongly suggests that the considered phenomena are related to the {\em imaginary} part of the conductivity $\sigma''_{xx}(\omega)$. 

\paragraph*{4.} What could this $\sigma''_{xx}$-related effect be? It is known that in the  inhomogeneous oscillating electromagnetic field electrons experience a nonlinear, time-independent {\em ponderomotive force}, see e.g. Refs. \cite{Boot57,Kibble66,Aamodt77,Krapchev79,Kaplan05}, proportional to the gradient of the squared electric field. Usually the ponderomotive forces are observed in the very intense laser fields. To the best of our knowledge, they have never been seen in the 2DEG systems. In the absence of scattering and the external magnetic field the ponderomotive force acting on a charged particle is  ${\bf F}_{pm}({\bf r})=-\nabla U_{pm}({\bf r})$ where 
\begin{equation} 
U_{pm}({\bf r})=\frac {e^2}{4m^\star \omega^2}\langle {\bf E}^2({\bf r},t)\rangle_t \equiv
\frac {\sigma''(\omega)}{4n_s\omega}\langle {\bf E}^2({\bf r},t)\rangle_t
\label{pondB0}
\end{equation}
is the ponderomotive potential and $\langle \dots \rangle_t$ means the averaging over time. The potential $U_{pm}({\bf r})$ is  proportional to the imaginary part of the dynamic Drude conductivity $\sigma''(\omega)$ per particle. In zero magnetic field the ponderomotive force always directed from the areas of the large field to the areas of the weaker field (independent of the charge of the particles).

\begin{figure}
\includegraphics[width=12cm]{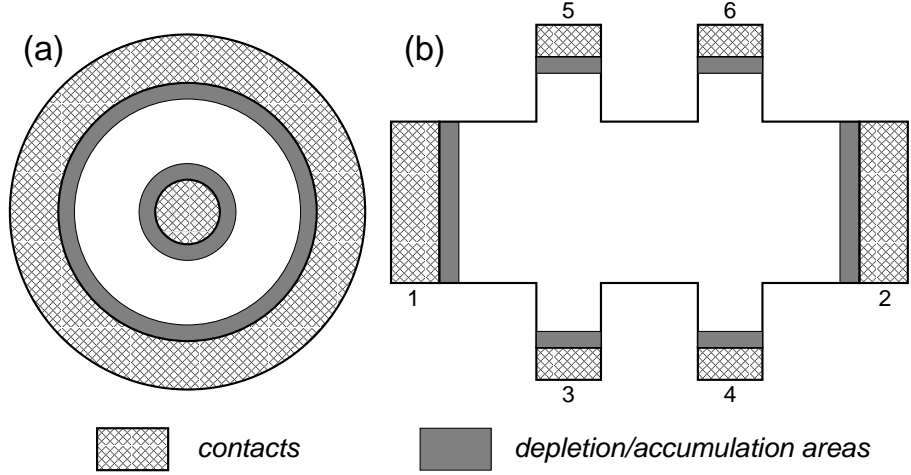}
\caption{\label{CorbHallgeom} The geometry of (a) the Corbino disk and (b) the Hall-bar sample under the intense microwave irradiation. The gray areas near the contacts show the microwave induced depletion/accumulation regions.}
\end{figure}

In the presence of the magnetic field ${\bf B}=(0,0,B)$, the linearly polarized electric field ${\bf E}({\bf r},t)= {\bf e}_x E_x(x)\cos\omega t$, as well as a weak scattering ($\gamma\ll\omega,\omega_c$), Eq. (\ref{pondB0}) is generalized as follows 
\begin{eqnarray}
&& U_{pm}(x)=\frac {\sigma_{xx}''(\omega)}{4n_s\omega}\langle  {\bf E}^2({\bf r},t)\rangle_t
\nonumber \\
&=& \frac{e^2E_x^2(x)}{8m^\star\omega_c}
\Bigg(
\frac{ \omega-\omega_c}{(\omega-\omega_c)^2+\gamma^2}
-
\frac{ \omega+\omega_c}{(\omega+\omega_c)^2+\gamma^2}
\Bigg),\label{PPlocal}
\end{eqnarray}
where 
\begin{equation} 
\sigma_{xx}(\omega)=\frac {in_se^2}{m^\star}\frac{\omega+i\gamma}{(\omega+i\gamma)^2-\omega_c^2}
\label{sigmaxx}
\end{equation}
is the diagonal dynamic Drude conductivity of the 2DEG [Eq. (\ref{PPlocal}) is derived below in Section \ref{TheoryNonlocal}, see also Ref. \cite{Aamodt77}]. Two points are very important here. First, the absolute value of the ponderomotive force $F_{pm}\propto \partial[E_x^2(x)]/\partial x$ dramatically grows near the contacts, Figure \ref{screen}(a,b). Second, in finite magnetic fields the force $F_{pm}$ {\em may change its direction}. If $\omega>\omega_c$, electrons are pushed away from the high-field areas (from the contacts), while at $\omega<\omega_c$ they are attracted to such areas. The density of electrons in the near-contact region will then be reduced or increased dependent on the sign of $\omega-\omega_c$. In other words, the microwaves form a depletion or an accumulation layer near the contacts (at $\omega>\omega_c$ and $\omega<\omega_c$ respectively), Figure \ref{CorbHallgeom}. 

\paragraph*{5. } The microwave induced near-contact depletion/accumulation regions influence the experimentally measured magnetotransport coefficients. Consider the most interesting case when the ponderomotive forces form a depletion layer and microwaves suppress the resistance/conductance of the sample. In the Corbino geometry, Figure \ref{CorbHallgeom}(a), all the applied voltage then drops on the depletion regions and one observes the vanishing {\em conductance} of the Corbino disks. The effect is mainly determined by the ratio ${\cal N}=n_s^c/n_s^0$ of the near-contact electron density $n_s^c$ to the bulk density $n_s^0$.

\begin{figure}
\begin{center}
\hfill \includegraphics[width=7.5cm]{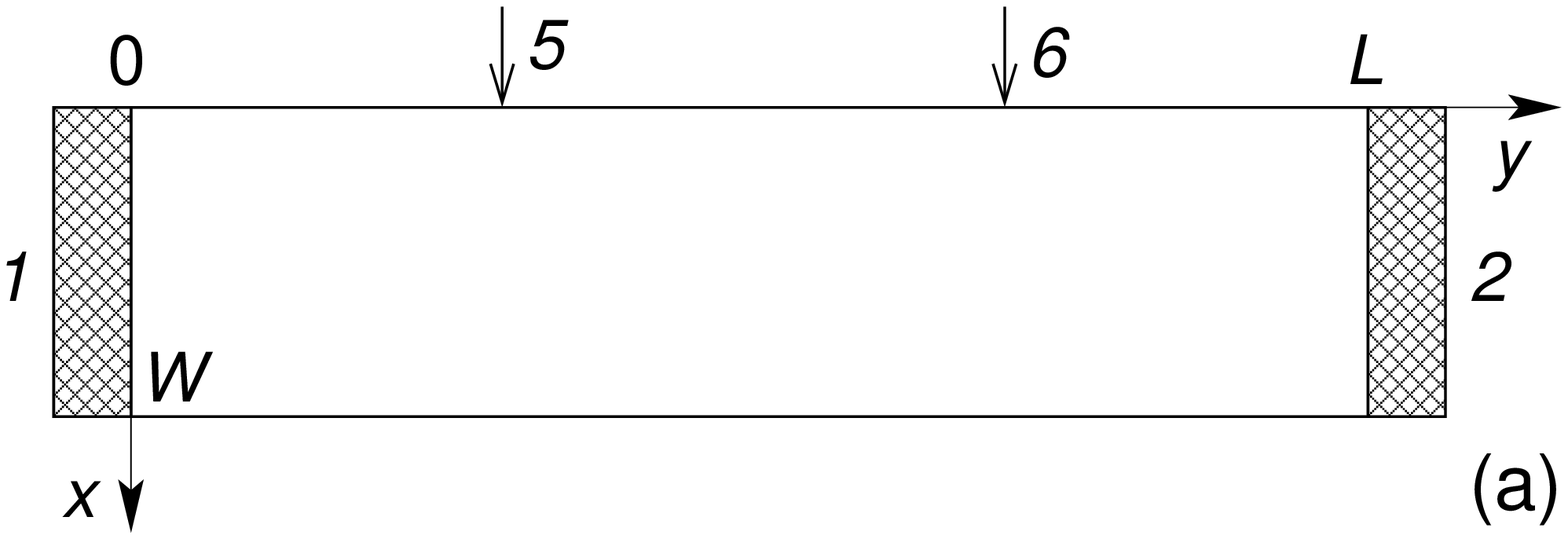}
\hfill 
\includegraphics[width=7.5cm]{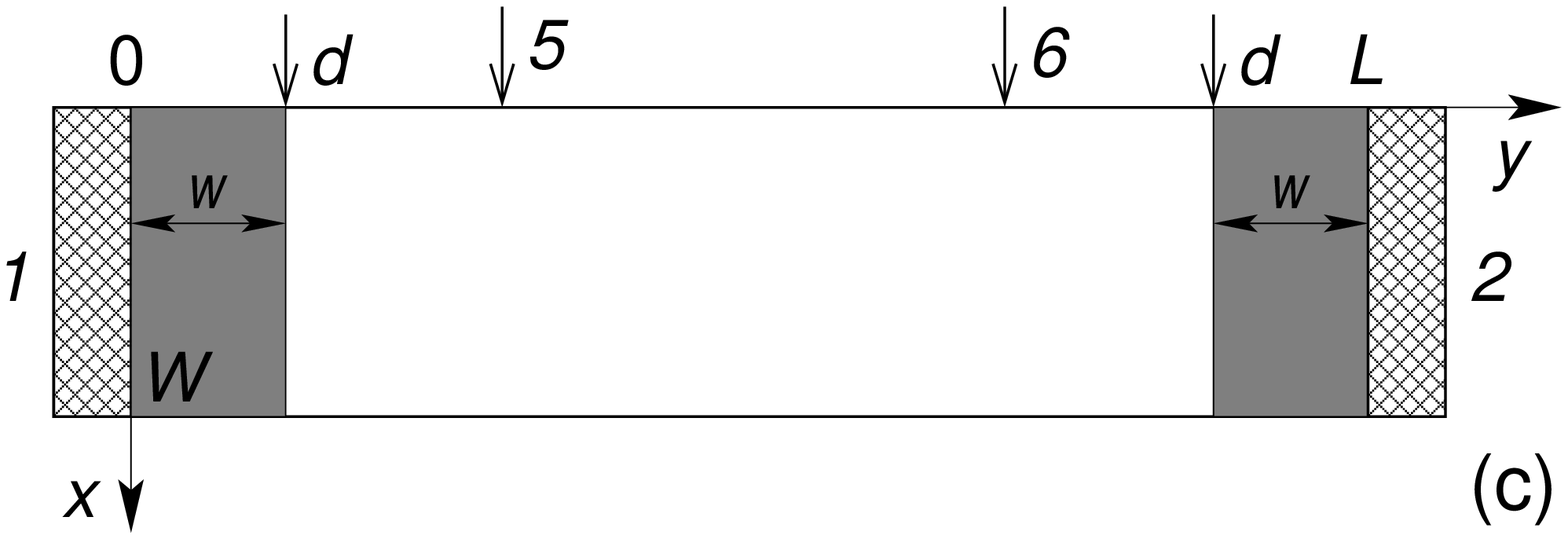} \hfill \\
\end{center}
\includegraphics[width=0.49\textwidth]{fig5b.eps}
\includegraphics[width=0.49\textwidth]{fig5d.eps}\\
\caption{\label{corners} (a) and (c): A rectangular sample with $\phi(x,0)=0$ (``west'' contact) and $\phi(x,L)=\phi_0$ (``east'' contact) (a) in the absence of microwaves and (c) irradiated by microwaves. The boundary conditions at the ``north'' and ``south'' sides of the rectangle are $j_x(0,y)=j_x(W,y)=0$. $W$ is the sample width in the $x$-direction, $L$ is the sample length in the $y$-direction; the arrows $5$ and $6$ show the position of the contacts; the arrows $d$ show the boundaries of the depletion layers. (b) and (d): The distribution of the dc electric potential on the ``north'' and ``south'' sides of the rectangle (b) in the absence of microwaves and (d) under the microwave irradiation. $U_{xy}$ and $U_{xx}$ schematically show the measured values of the Hall and diagonal voltages.}
\end{figure}

In the Hall-bar geometry, Figure \ref{CorbHallgeom}(b), similarly, the formation of the depletion layers near the current contacts 1 and 2 suppresses the measured voltage between the side contacts $U_{xx}=U_{34}\simeq U_{56}$. To show this,  consider first a uniform Hall-bar sample not irradiated by microwaves, Figure \ref{corners}(a). It is known \cite{Rendell81,Thouless85} that the flowing dc current produces a strongly inhomogeneous distribution of the dc electrical potential $\phi(x,y)$ in the Hall bar, with power-law singularities at the diagonally opposite corners of the rectangular sample. For example, near the corner $(x,y)=(0,0)$ 
\begin{equation} 
\phi(x,y)\propto r^\eta\sin\eta\theta,
\label{singular_pot}
\end{equation}
where 
\begin{equation}  
\eta=\frac{2\sigma_{xx}}{\pi\sigma_{xy}}\ll 1 ,
\end{equation}
$(r,\theta)$ are the polar coordinates of the point $(x,y)$ and we assume that $\sigma_{xy}/\sigma_{xx}\gg 1$. Figure \ref{corners}(b) qualitatively illustrates the behavior of the potential at the ``north'' and ``south'' sides of the Hall bar, $\phi(0,y)$ and $\phi(W,y)$ (we have used $\sigma_{xy}/\sigma_{xx}=5$ in this Figure; in the experiment this ratio is much higher, e.g. $\sigma_{xy}/\sigma_{xx}\simeq 250$ in Ref. \cite{Mani02}, therefore all the features discussed here are much stronger in the real experiments). The measured values of the Hall and diagonal voltages $U_{xy}$ and $U_{xx}$ in the absence of microwaves are also shown there. 

Now assume that the sample is irradiated and the depletion regions at $y<w$ and $y>L-w$ are formed near the contacts 1 and 2 [gray areas in Figure \ref{corners}(c)]. Further, assume for simplicity that the density of electrons in the depletion regions and in the bulk of the sample are constant and equal $n_s^c$ and $n_s^0$ respectively. The tangential electric current $j_y$ near the ``north'' and ``south'' sides of the rectangle must be continuous at the boundaries of the depletion regions, i.e. $j_y(0,w-0)=j_y(0,w+0)$, $j_y(W,L-w-0)=j_y(W,L-w+0)$. In addition, the current $j_x$ should vanish in the corresponding boundary points, $j_x(0,w)=j_x(W,L-w)=0$. This leads to a jump of the tangential electric field at the boundaries of the depletion regions,
\begin{equation} 
\left(\frac{\partial\phi}{\partial y}\right)_0= \frac{\rho_{xx}^0}{\rho_{xx}^c}\left(\frac{\partial\phi}{\partial y}\right)_c =\frac{n_s^c}{n_s^0}\left(\frac{\partial\phi}{\partial y}\right)_c={\cal N}\left(\frac{\partial\phi}{\partial y}\right)_c.\label{slopechange}
\end{equation}
As seen from Figure \ref{corners}(d), the discontinuity of the field (\ref{slopechange}) leads to a reduction of the measured voltage $U_{xx}$, proportional to the density factor ${\cal N}$. The measured diagonal photoresistance $R_{xx}$ can then be written as a product of two terms, 
\begin{equation}
R_{xx}\simeq R_{xx}^b{\cal N}. \label{Rxx}
\end{equation}
The first factor $R_{xx}^b$  is the photoresistance of the uniform sample (the bulk contribution). It is due to the resonant absorption of microwaves in the bulk and has a Lorentzian shape $R_{xx}^b\sim  1/[(\omega-\omega_{mp})^2+\gamma^2]$ with the absorption maximum at the magnetoplasmon frequency (\ref{magnetopl}), Figure \ref{compar}(a). The second factor ${\cal N}$ describes the change of the measured photoresistance due to the near-contact microwave-induced inhomogeneity (the contact contribution). In the next two paragraphs we show that the density factor ${\cal N}$ is almost always very close to unity; it may substantially differ from ${\cal N}\approx 1$ only in the very-high-mobility samples, i.e. under the conditions of the MIZRS experiments. In contrast, the magnetoplasmon resonance from the bulk contribution $R_{xx}^b$ is always present in the measured signal. As a result, in the low-mobility samples one observes  only the magnetoplasmon-resonance response \cite{Vasiliadou93}. In the MIRO experiments \cite{Zudov01} (relatively weak resistance oscillations, moderate-mobility samples) one sees both the magnetoplasmon resonance and the $\omega_c$-related oscillations. Finally, in the extremely clean samples only the giant $\omega_c$-oscillations are observed since the weak $\omega_{mp}$ resonance is hidden under the huge oscillations of $R_{xx}$. This explains one of the main puzzles of the MIZRS/MIRO effects.

Figure \ref{corners}(d) also shows that the measured variations of the Hall voltage $U_{xy}$ are very small as compared to $U_{xx}$ if $\sigma_{xy}/\sigma_{xx}\gg 1$. This again agrees with the MIZRS experiments.

\paragraph*{6.} Now consider the density parameter ${\cal N}=n_s^c/n_s^0$ quantitatively. In the absence of microwaves the density $n_s^0$ in the uniform sample is
\begin{equation} 
n_s^0=\frac{m^\star T}{ \pi\hbar^2}F\left(\frac {\zeta_0} T\right),
\end{equation}
where 
\begin{equation}
F(z)=\int_0^\infty \frac{dx}{1+\exp(x-z)}\approx \Bigg\{
\begin{array}{ll}
z,\ & z>0,\ |z|\gg 1,\\
e^{z},\ & z<0,\ |z|\gg 1,
\end{array}
\label{Fermif-n}
\end{equation}
is the Fermi integral in the 2D case and $\zeta_0=E_F$ is the chemical potential (the Fermi energy). In the Fermi gas $\zeta_0\gg T$ and $n_s^0=m^\star \zeta_0/ \pi\hbar^2$. In the presence of the microwave induced ponderomotive potential $U_{pm}(x)$, Eq. (\ref{PPlocal}), the density of electrons becomes a function of $x$, 
\begin{equation}
n_s(x) =\frac{m^\star T}{ \pi\hbar^2}F\left(\frac {\zeta -U_{pm}(x)}{T}\right).
\label{n_s}
\end{equation}
The chemical potential $\zeta$ here may, in general, differ from $\zeta_0$, but, since the 2DEG is always connected to the contact reservoirs in the discussed experiments we will assume that $\zeta=\zeta_0=\pi\hbar^2n_s^0/m^\star$. Now, rewrite the density (\ref{n_s}) in the form
\begin{eqnarray}
\frac{ n_s(x)}{n_s^0}&=&
\frac T{\zeta_0} F\left(\frac {\zeta_0} T \left[1-\frac {U_{pm}(x)}{\zeta_0}\right]\right)\label{n_s1} \\ &=&
\frac T{\zeta_0} F\left(\frac {\zeta_0} T \Big[1-{\cal P}{\cal F}^2(x){\cal B}_1(\Omega_c,\Gamma)\Big]\right),\label{n_s1b}
\end{eqnarray}
where the factor
\begin{equation} 
{\cal P}= \frac 18\left(\frac{eE_0v_F}{\omega E_F}\right)^2=\frac{\pi e^2 P}{m^\star c \omega^2E_F} \label{PowerFactor}
\end{equation}
is proportional to the power $P=cE_0^2/4\pi$ (per unit area) of the incident radiation, $v_F$ is the Fermi velocity, 
\begin{equation} 
{\cal F}(x)=\frac{E_x(x)}{E_0}
\end{equation} 
is the electric field really acting on the electrons, normalized by the external field amplitude $E_0$, and 
\begin{equation}
{\cal B}_k(\Omega_c,\Gamma)=
\frac {\omega^2}{2\omega_c}
\Bigg(
\frac{ \omega-k\omega_c}{(\omega-k\omega_c)^2+\gamma^2}
-
\frac{ \omega+k\omega_c}{(\omega+k\omega_c)^2+\gamma^2}
\Bigg)
\label{Bfactor}
\end{equation}
is a factor dependent on the dimensionless magnetic field  $\Omega_c=\omega_c/\omega=eB/m^\star c\omega$ and the scattering rate  $\Gamma=\gamma/\omega$ [in the simplified expressions (\ref{n_s1b}), (\ref{PPlocal}) only the factor ${\cal B}_1$ is used; more general formulas with all ${\cal B}_k$, $k=1,2,3,\dots$, will appear later, see Eq. (\ref{PMfinal})]. 

\begin{figure}
\includegraphics[width=12cm]{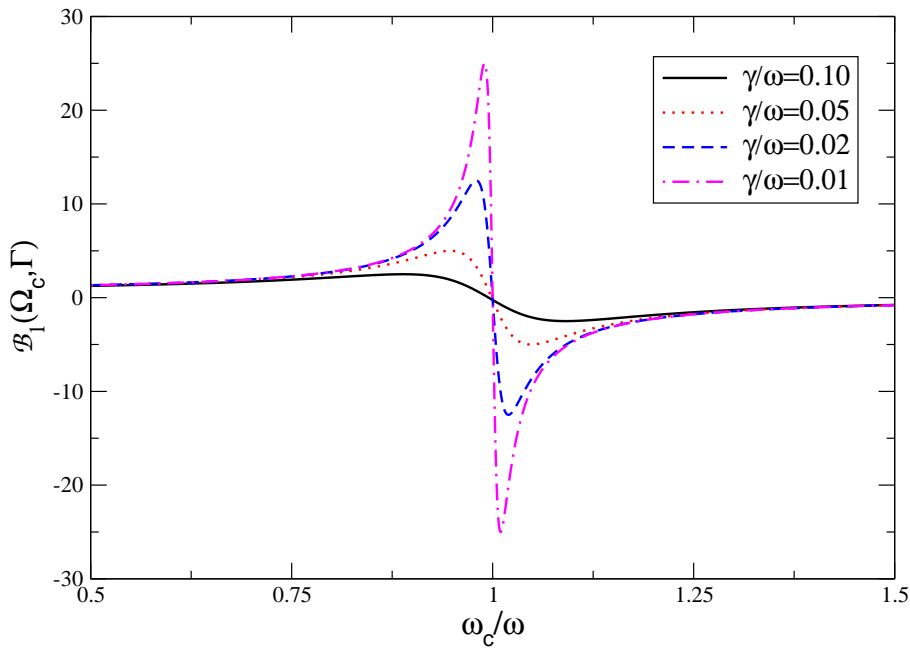}
\caption{\label{bfactor} (Color online) The $B$-dependent factor ${\cal B}_1(\Omega_c,\Gamma)$, Eq. (\ref{Bfactor}). At the mobility $\mu=2\times 10^7$ cm$^2$/Vs and the frequency $100$ GHz the value of $\gamma/\omega$ is $\gamma/\omega\simeq 2\times 10^{-3}$ (even smaller than shown in this Figure), therefore the resonance will be even sharper. }
\end{figure}

\paragraph*{7.} Equation (\ref{n_s1b}) allows one to understand (i) why the ponderomotive effects have not been seen in the earlier experiments, (ii) why the MIRO/ZRS phenomenon can be observed only in the ultraclean samples, (iii) how the discussed effects depend on the microwave power and temperature and (iv) which factors favor and impede the observation of the MIZRS effect. The ponderomotive forces noticeably change the 2D electron density in the near-contact regions only if the correction 
\begin{equation} 
\frac{U_{pm}(x)}{\zeta_0}={\cal P}{\cal F}^2(x){\cal B}_1(\Omega_c,\Gamma)\label{correction}
\end{equation} 
in the square brackets in (\ref{n_s1b}) is not negligible as compared to unity. The basic factor in (\ref{correction}) is the power parameter ${\cal P}\propto (eE_0v_F/\omega E_F)^2$, Eq. (\ref{PowerFactor}). The factor $(eE_0v_F/\omega E_F)$ here is the ratio of the energy which electrons get from the external field during one oscillation period ($\sim eE_0v_F/\omega$) to their average energy ($E_F$). Usually this parameter is extremely small: taking for instance the typical MIRO/ZRS experimental values -- $P\simeq 1$ mW/cm$^2$, $f=\omega/2\pi\simeq 100$ GHz, $m^\star\approx 0.067 m_0$ and $n_s^0\simeq 3\times 10^{11}$ cm$^{-2}$ -- we get  ${\cal P}\simeq 1.3\times 10^{-6}$. Therefore the nonlinear electromagnetic phenomena can be seen only in the very intense  fields and therefore the ponderomotive forces have not been observed in the 2DEG systems earlier. In the MIZRS experiments, however, the small parameter ${\cal P}$ in (\ref{correction}) is multiplied by {\em two very large factors}. First, the field parameter ${\cal F}^2(x)$  describes the giant growth of the electric field in the  near-contact areas, Figure \ref{screen}(a,b). Second, the $B$-dependent factor ${\cal B}_1(\Omega_c,\Gamma)$ becomes extremely large near the cyclotron resonance $\omega\simeq\omega_c$ in the very-high-mobility samples, Figure \ref{bfactor}. 

How the MIRO and MIZRS depend on the microwave power and temperature? As seen from Eqs. (\ref{n_s1b}) and (\ref{Fermif-n}) there exist two different regimes. If the microwave power is not very strong, $|U_{pm}|/\zeta_0\lesssim 1$, the density factor varies linearly with the power, 
$${\cal N}\simeq 1-{\cal P}{\cal F}_c^2{\cal B}_1(\Omega_c,\Gamma),$$
${\cal F}_c=E_c/E_0\gg 1$. This is the regime of MIRO -- the microwave induced resistance oscillations \cite{Zudov01,Ye01}. The temperature dependence of the measured signal in this regime is weak. If the microwave power is so strong that the parameter (\ref{correction}) exceeds unity, $U_{pm}/\zeta_0\gtrsim 1$, the density of electrons in the depletion regions becomes exponentially small (the Fermi gas becomes the Boltzmann one) and one gets into the regime of MIZRS -- the microwave induced ``zero-resistance'' states. The $P$ and $T$ dependencies in the MIZRS regime are described by the Arrhenius-type law 
$$ {\cal N}\propto \exp\left(-\frac{\pi e^2 P{\cal F}_c^2}{m^\star c \omega^2T}{\cal B}_1(\Omega_c,\Gamma)\right).$$
Such behavior of the signal as a function of power and temperature agrees with the experiments, see e.g. Refs. \cite{Mani02,Zudov03,Willett04,Mani04}. 

Which factors favor and which impede the observation of the ``zero-resistance states''? The transition from the MIRO to the MIZRS regime is the case when the parameter (\ref{correction}) becomes bigger than unity. Taking the maxima of ${\cal F}(x)\simeq {\cal F}_c= E_c/E_0$ (at the contacts) and of ${\cal B}_1(\Omega_c,\Gamma)$ (at $\omega-\omega_c=\gamma$) we get the MIZRS observability condition in the form 
\begin{equation} 
\frac{e^2}{\hbar c } \frac{P}{ \hbar \omega n_s^0\gamma}\frac{E_c^2}{E_0^2} = \frac{em^\star }{\hbar^2 c }\times  \left(\frac{P\mu}{  \omega  n_s^0 }
{\cal F}_c^2 \right)\gtrsim 1.
\label{MIZRSobservability}
\end{equation}   
The higher the radiation power $P$ and the electron mobility $\mu$, the easier is it to observe MIZRS. These dependencies have been clear from the very first MIZRS experiments. On the other hand, the high electron density $n_s^0$ and radiation frequency $\omega$ impede the observation of MIZRS. The influence of $n_s^0$ on the MIZRS effect has not been systematically  studied, but the suppression of MIZRS at high frequencies has been recently reported by Studenikin et al. \cite{Studenikin07}. The $\omega$-dependence thus also agrees with the experimental facts. 

One more important factor influencing the MIZRS observability is the shape and the quality of the contacts. The ratio ${\cal F}_c$ of the near-contact electric field to the external one will be larger if the contact is closer to the ideal conditions (the infinite conductivity $\sigma_c$ and the vanishing thickness in the $z$-direction). There have been no systematic study of the role of the contacts in the microwave experiments on the 2DEG, but if the contacts, due to some reasons, turned out to be ``more ideal'' in the discussed sense, the observation of MIZRS could become possible even in samples with a moderate mobility. This may explain the experiment \cite{Bykov06} in which the giant magnetoresistance oscillations and zero-resistance states have been observed at $\mu\lesssim 10^6$ cm$^2$/Vs. 

The ``negative resistance'' observations of Ref. \cite{Willett04} and the influence of the parallel magnetic fields \cite{Yang06} on the MIZRS can be also explained by the influence of  contacts. In the first case one should take into account that the depletion regions are formed not only near the current contacts 1 and 2 but also near the side contacts 3, 4, 5 and 6, Figure \ref{CorbHallgeom}(b). Since the local properties of the contacts can be different, the dc field near the contacts is distorted and one can measure, in principle, slightly negative values of $U_{xx}$.  

The second effect \cite{Yang06} -- the suppression of  MIZRS by the parallel magnetic fields  $B_\parallel\sim 1$ T -- seems to be completely unbelievable if to think about the influence of $B_\parallel$ on the properties of the 2DEG. This becomes, however, quite reasonable, if to assume that $B_\parallel$ modifies the contact properties. In our model the contacts are infinitely thin and have the $B$- and $\omega$-independent conductivity $\sigma_c$. In reality the contacts are three-dimensional and their conductivity tensor may be quite sensitive to the parallel magnetic fields of order of $ 1$ T (at $B\simeq 1$ T, $m^\star\simeq m_0$ and  $f\simeq  30$ GHz the microwave and the cyclotron frequencies are equal). The suppression of  MIZRS by the parallel magnetic fields can then be explained by the influence of  $B_\parallel$ on the contact factor ${\cal F}_c$.

\paragraph*{8.} So far we have discussed the microwave induced phenomena in the 2DEG using the simplified formula (\ref{PPlocal}) for the ponderomotive potential. This approach is valid under the condition 
\begin{equation} 
\frac{R_c}{E_x(x)}\frac{dE_x(x)}{dx}\ll 1,\label{quasi-loc-cond}
\end{equation}
when the inhomogeneity of the electric field is taken into account in the lowest order (the quasi-local approximation). Eq. (\ref{PPlocal}) explains the resonance behavior of the density factor ${\cal N}$  near the fundamental cyclotron harmonics $\omega\simeq \omega_c$ but does not describe the resonances around  $\omega\simeq k\omega_c$. In order to explain the microwave induced oscillations at higher cyclotron harmonics, one needs a more general, {\em nonlocal} theory of the ponderomotive forces. 

Such a theory is developed in Section \ref{TheoryNonlocal}. It is shown there that the general expression for the ponderomotive potential has the form
\begin{equation}
U_{pm}(x)=
\frac {e^2}{4m^\star\omega^2} 
\sum_{k=1}^\infty 
\Big[\epsilon_{k-1}^2 (x)-\epsilon_{k+1}^2 (x)\Big] 
{\cal B}_k(\Omega_c,\Gamma)
,\label{PMfinal}
\end{equation}
where the factors ${\cal B}_k(\Omega_c,\Gamma)$ are defined in (\ref{Bfactor}) and 
\begin{equation} 
\epsilon_k(x)=\frac 1{\pi}\int_{0}^\pi E_x(x+R_c\cos\xi)\cos k\xi d\xi.\label{epsdef}
\end{equation} 
Eq. (\ref{PMfinal}) contains all the cyclotron harmonics, with the amplitude of the $k$-th term determined by the $x$-dependence of the electric field $E_x(x)$, Eq. (\ref{epsdef}). In the special case of $R_c\to 0$ [the field slowly varies at the cyclotron radius scale, Eq. (\ref{quasi-loc-cond})] Eq. (\ref{epsdef}) gives $\epsilon_k(x)\approx  E_x(x)\delta_{k0}$. Then only the first term with $k=1$ remains in the sum of Eq. (\ref{PMfinal}) and we obtain Eq. (\ref{PPlocal}). 
In the general, essentially nonlocal, case the potential (\ref{PMfinal}) should be substituted into Eq. (\ref{n_s1}). Then one gets the density factor ${\cal N}$ with oscillations at all cyclotron harmonics ($x_c\lesssim R_c$),
\begin{equation} 
{\cal N}\simeq \frac T{\zeta_0} F\left(\frac {\zeta_0} T \left[1-\frac {e^2}{4m^\star\omega^2\zeta_0} 
\sum_{k=1}^\infty 
\Big[\epsilon_{k-1}^2 (x_c)
-\epsilon_{k+1}^2 (x_c)
\Big] 
{\cal B}_k(\Omega_c,\Gamma)\right]\right).\label{calNgeneral}
\end{equation} 

\begin{figure}
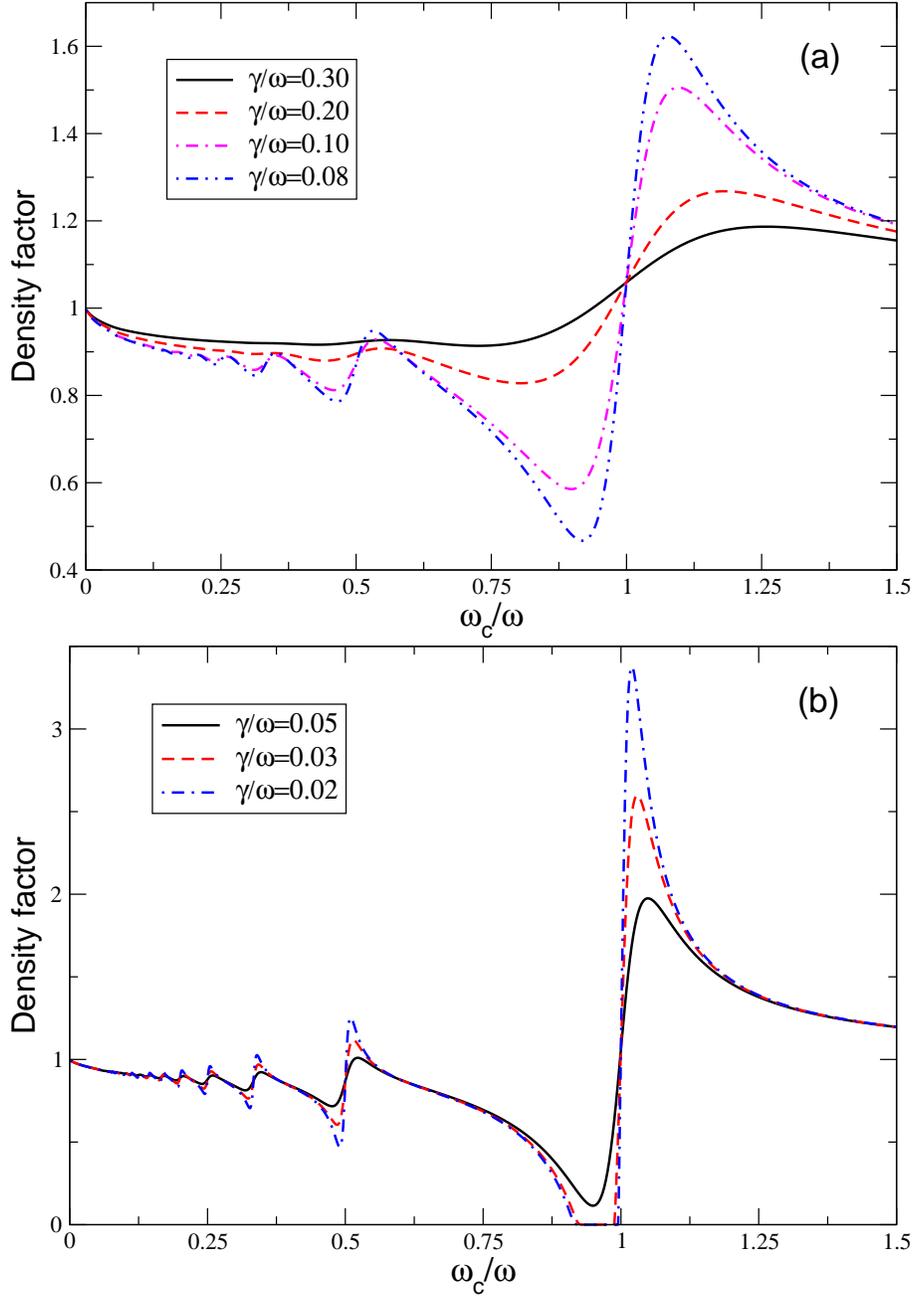

\includegraphics[width=12cm]{fig7a.eps}\\
\includegraphics[width=12cm]{fig7b.eps}
\caption{\label{fig:densfactor}  (Color online) The  density factor ${\cal N}$, Eq. (\ref{DensFactorModel}), as a function of the magnetic field at different mobilities of the 2D electrons: (a) the microwave induced oscillations in the moderate mobility regime (notice the vertical axis scale); (b) the formation of the ``zero resistance'' states in the very high mobility regime. Other parameters are: ${\cal P}_c={\cal PF}_c^2=1$, $T/\zeta_0=0.02$, $v_F/\omega l=8$. }
\end{figure}

\begin{figure}
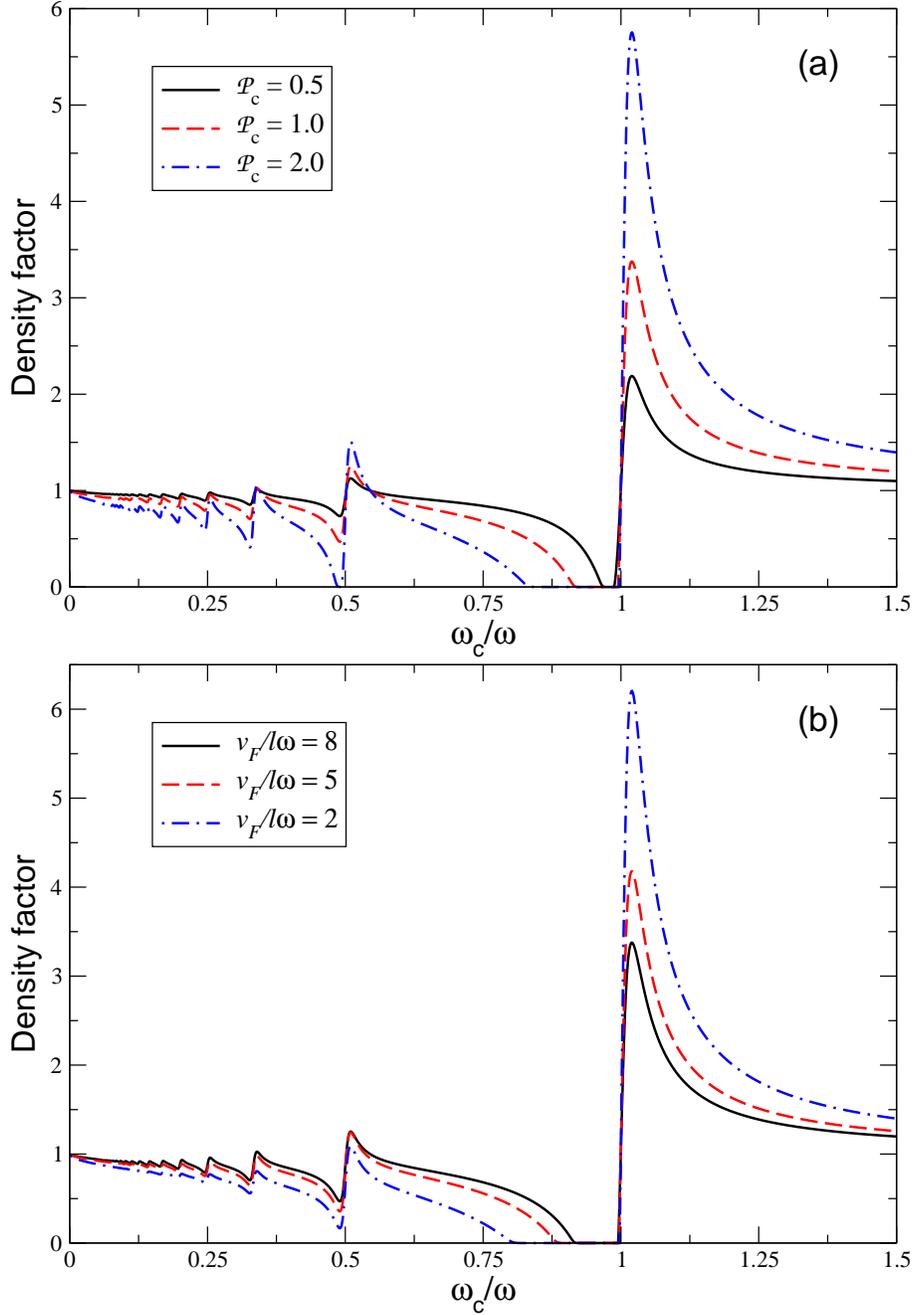

\includegraphics[width=12cm]{fig8a.eps}\\
\includegraphics[width=12cm]{fig8b.eps}
\caption{\label{fig:PowerNonloc}  (Color online) The influence of (a) the power parameter ${\cal P}_c={\cal PF}_c^2$ and (b) the nonlocality parameter $v_F/\omega l$ on the MIZRS. Other parameters used: $T/\zeta_0=0.02$, $\gamma/\omega =0.02$, as well as (a) ${\cal P}_c=1$ and (b) $v_F/\omega l=8$. }
\end{figure}

\paragraph*{9. } Equation (\ref{calNgeneral}) gives a general formula for the near-contact contribution to the measured microwave induced photoresistance of the 2DEG. It depends on the behavior of the electric field near the contacts. To make our results more specific we consider the following model for the  electric field distribution
\begin{equation} 
E_x(x)=E_c\sqrt{\frac l{l+x}}.\label{sqrfield}
\end{equation}
This model describes the square-root divergency of the field near the contact which is cut off at a length $l$, $l\ll R_c$ (in a real sample $l$ can be estimated as a distance between the contact and the 2D electron layer, i.e. $l\simeq 0.1$ $\mu$m). Calculating the coefficients $\epsilon_k(x)$ for the field (\ref{sqrfield}) we  get the density factor ${\cal N}$ in the form (see details in Section \ref{TheoryNonlocal})
\begin{equation} 
{\cal N}=\frac T{\zeta_0} F\left(\frac {\zeta_0} T \left[1-{\cal P}{\cal F}_c^2 
\sum_{k=1}^\infty 
T_{k}\left(\frac{v_F}{\omega l}\frac 1{\Omega_c}\right)
{\cal B}_k(\Omega_c,\Gamma)\right]\right),\label{DensFactorModel}
\end{equation}
where the functions $T_k(z)$ are defined below in Section \ref{secB} [see Eq. (\ref{eq-Tk})]. One sees that the factor ${\cal N}$ depends on five  parameters: 
\begin{itemize}
\item the dimensionless magnetic field $$\Omega_c=\frac{\omega_c}\omega=\frac{eB}{m^\star c\omega},$$ 
\item 
the dimensionless scattering rate (the inverse mobility) 
$$\Gamma=\frac \gamma\omega=\frac e{m^\star\mu\omega},$$ 
\item 
the power parameter (proportional to the squared {\em contact} field electric $E_c$) $${\cal P}_c={\cal PF}_c^2=\frac 18\left(\frac{eE_cv_F}{\omega E_F}\right)^2,$$ 
\item 
the dimensionless temperature $T/\zeta_0$, and 
\item 
the nonlocality parameter $v_F/\omega l$. 
\end{itemize}
Figures \ref{fig:densfactor} and \ref{fig:PowerNonloc} exhibit the influence of the different parameters of the problem on the density factor ${\cal N}$ and hence, on the observed photoresistance $R_{xx}$, Eq. (\ref{Rxx}).  In Figure \ref{fig:densfactor}(a) one sees the $B$ dependencies of the measured signal at low mobilities, i.e. in the regime of MIRO \cite{Zudov01,Ye01}. One observes how the almost constant value of the measured signal is transformed into the oscillating behavior. The calculated curves quite accurately reproduce those measured in the first MIRO experiments \cite{Zudov01,Ye01}. In Figure \ref{fig:densfactor}(b) the mobilities are higher and one gets into the ``zero resistance'' regime. A large number of higher harmonics is observed. Figure \ref{fig:PowerNonloc}(a) shows that the growth of the microwave power increases the width of the ``zero resistance'' regions and the amplitude of higher harmonics. The influence of the nonlocality parameter, Figure \ref{fig:PowerNonloc}(b), is more complicated. Its reduction leads to smaller oscillation amplitudes at higher $k$ but increases the width of the ``zero resistance'' region at $k=1$. The overall agreement of the presented analytical theory with the  experimental data \cite{Mani02,Zudov03,Dorozhkin03,Yang03,Kovalev03,Willett04,Zudov04,Mani04,Mani04b,Mani04c,Studenikin04,Du04,Smet05,Studenikin05,Mani05,Yang06,Zudov06a,Zudov06b,Studenikin07,Andreev08,Hatke09}  is evident. 

The development of the theory of MIRO/ZRS is thereby completed.

\section{Nonlocal theory of the Ponderomotive forces\label{TheoryNonlocal}}

\subsection{General approach}

Consider the classical motion of a 2D electron in the uniform magnetic field ${\bf B}=(0,0,B)$ and the nonuniform oscillating electric field ${\bf E}({\bf r},t)= {\bf e}_x E_x(x)\cos\omega t$. The equations of motion have the form 
\begin{equation} 
{\bf \dot r=v},\label{rdot}
\end{equation} 
\begin{equation} 
m^\star{\bf \dot v}=-\frac ec {\bf v}\times {\bf B} -\gamma m^\star {\bf v}+F_x(x,t){\bf e}_x  ,\label{vdot}
\end{equation}
where $F_x(x,t)=-eE_x(x)\cos\omega t$ and the scattering of electrons is taken into account by the friction term $-\gamma {\bf v}$. The scattering is assumed to be small, $\gamma/\omega\ll 1$. 

In the zeroth order in the electric field amplitude $E_x$ the equations (\ref{rdot})--(\ref{vdot}) have the solution
\begin{equation} 
{\bf v}^{(0)}(t)= v_0\left(
\begin{array}{r}
\cos(\omega_ct+\phi) \\
\sin(\omega_ct+\phi) \\
\end{array}
\right), 
\end{equation} 
\begin{equation} 
{\bf r}^{(0)}(t)=\left(
\begin{array}{r}
x_0  \\
y_0 \\
\end{array}
\right)+ R_c 
\left(
\begin{array}{r}
\sin(\omega_ct+\phi) \\
-\cos(\omega_ct+\phi) \\
\end{array}
\right),\label{rsol}
\end{equation} 
i.e. the particle rotates around the point $(x_0,y_0)$ with the cyclotron frequency; here $x_0$, $y_0$, $v_0$ and $\phi$ are integration constants and $R_c= v_0/\omega_c$ is the cyclotron radius. In the final formulas for the 2DEG system one can estimate $R_c$ as $R_c=v_F/\omega_c$, where $v_F$ is the Fermi velocity. Since the electric field is strongly inhomogeneous on the $R_c$-scale, the particle experiences different forces at the different parts of its trajectory and the net average force acting on it turns out to be nonzero. This time-independent force, which appears in the second order in the field amplitude $E_x$, is the ponderomotive force $F_{pm}(x_0)$  we are looking for. In order to find it we first search for the solution of Eqs. (\ref{rdot})--(\ref{vdot}) in the first order in $E_x$.  Substituting $x^{(0)}(t)$ from (\ref{rsol}) into the force $F_x$ we get  
\begin{equation} 
F_x = - e E_x(x_0+R_c\cos\xi)\cos\omega t ,\label{vdot1}
\end{equation}
where $\xi=\omega_ct+\phi-\pi/2$. The function $E_x(x_0+R_c\cos\xi)$ is a periodic function of $\xi$ with the period $2\pi$. Expanding it in the Fourier series
\begin{equation} 
E_x(x_0+R_c\cos\xi)=\sum_{k=-\infty}^\infty \epsilon_k(x_0) e^{ik\xi},\label{Eexpan}
\end{equation} 
we present $F_x$ as a sum of an infinite number of harmonics with the frequencies $\pm \omega +k\omega_c$. The first-order correction to the coordinate $x(t)$ then reads 
\begin{equation} 
x^{(1)}(t)=
\sum_{k=-\infty}^\infty 
\frac{A_ke^{i(\omega+k\omega_c)t}}{(\omega+k\omega_c-i\gamma)^2-\omega_c^2} 
+(\omega\to -\omega),\label{x1(t)}
\end{equation}
where 
\begin{equation} 
A_k=\frac {e}{2m^\star} e^{ik(\phi-\pi/2)} \epsilon_k(x_0)
\label{Pk1}
\end{equation}
and the  coefficients $\epsilon_k(x)$ in the Fourier expansion (\ref{Eexpan}) are  defined  in Eq. (\ref{epsdef}).

In the next order in $E_x$ we obtain
\begin{eqnarray} 
 F_x \approx
-e \cos\omega t \left( E_x[x^{(0)}(t)]
+ E'_x[x^{(0)}(t)]x^{(1)}(t) \right),\label{vdot5}
\end{eqnarray} 
where $E_x'(x)=\partial E_x(x)/\partial x$. The first term in Eq. (\ref{vdot5}) is the first-order force (\ref{vdot1}) which contains only the oscillating terms. The time-averaging of the second term should give the ponderomotive force. To calculate it we differentiate Eq. (\ref{Eexpan}) with respect to $x_0$ to get the Fourier expansion of the derivative $E'_x(x_0+ R_c\cos\xi)$,
\begin{equation} 
E'_x(x_0+R_c\cos\xi)=\sum_{k=-\infty}^\infty \frac{\partial \epsilon_k (x_0)}{\partial x_0}e^{ik\xi}.\label{E'expan}
\end{equation}  
Then we substitute (\ref{E'expan}) and (\ref{x1(t)}) into (\ref{vdot5}), average the resulting expression over time and after some algebraic transformations finally get the ponderomotive force in the conventional form $F_{pm}(x_0)=-\partial U_{pm}(x_0)/\partial x_0$ with the potential  (\ref{PMfinal}). 

\subsection{Ponderomotive potential for the model of Eq. (\ref{sqrfield})\label{secB}}

\begin{figure}
\includegraphics[width=12cm]{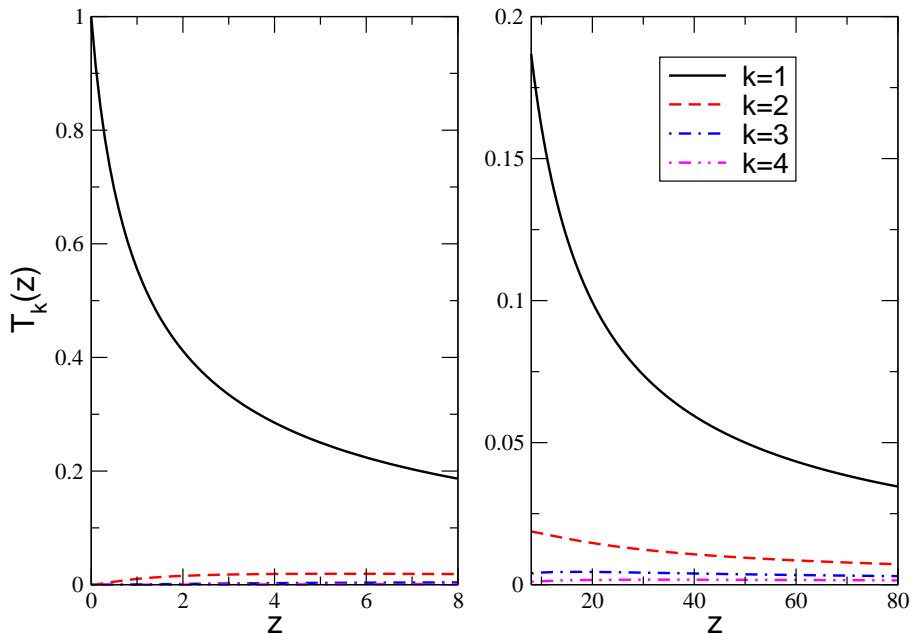}
\caption{\label{fig-Tk}  (Color online) The functions $T_k(z)$ defined in Eq. (\ref{eq-Tk}). }
\end{figure}

If the near-contact microwave electric field is described by the model expression  (\ref{sqrfield}) the coefficients $\epsilon_k$, Eq.  (\ref{epsdef}),  are written as 
\begin{equation} 
\epsilon_k(x)=
E_cS_k\left(1+\frac xl,\frac{R_c}l\right).\label{epsK}
\end{equation} 
Here the function
\begin{equation} 
S_k(a,b)=
\frac 1{\pi}\int_{0}^\pi \frac {\cos kxdx}{\sqrt{a+b\cos x}}
\end{equation} 
can be expressed \cite{PrudnikovI} in terms of the associated Legendre functions of the first kind $P_\nu^k(x)$,
\begin{equation} 
S_k(a,b)=
\frac{(a^2-b^2)^{-1/4}}{(1/2)_k}P_{-1/2}^{k}\left(\frac{a} {\sqrt{a^2-b^2}}\right),\ a>|b|.
\end{equation} 
Substituting (\ref{epsK}) into the general formula (\ref{PMfinal}), we get the ponderomotive potential in the form
\begin{equation}
U_{pm}(x)=
\frac {e^2E_c^2}{4m^\star\omega^2 } 
\sum_{k=1}^\infty 
R_{k}\left(1+\frac xl,\frac{R_c}l\right)
{\cal B}_k(\Omega_c,\Gamma)
,\label{PMsqrtmodel}
\end{equation}
where
$
R_k(a,b)=
S_{k-1}^{2}\left(a,b\right)
- 
S_{k+1}^{2}\left(a,b\right)
$. 
Evaluating the potential $U_{pm}(x)$ near the edge of the 2DEG we substitute $x\approx R_c$ in Eq. (\ref{PMsqrtmodel}) and get for $U_{pm}^c\equiv U_{pm}(x\sim R_c)$
\begin{equation}
U_{pm}^c\simeq 
\frac {e^2E_c^2}{4m^\star\omega^2 } 
\sum_{k=1}^\infty 
T_{k}\left(\frac {R_c}l\right)
{\cal B}_k(\Omega_c,\Gamma)
,\label{PMedge}
\end{equation}
where 
\begin{equation} 
T_k(z)=
S_{k-1}^{2}\left(1+z,z\right)
- 
S_{k+1}^{2}\left(1+z,z\right).\label{eq-Tk}
\end{equation} 
The functions $T_k(z)$ are plotted in Figure \ref{fig-Tk}.

Equation (\ref{PMedge}), together with (\ref{n_s1}), gives the expression (\ref{DensFactorModel}) for the density factor ${\cal N}$. 

\section{Discussion and Conclusions}

The  microwave induced giant oscillations of the magnetoresistance  are thus caused by the nonlinear ponderomotive forces which arise in the near contact areas. The ponderomotive phenomena are well known in the plasma physics (they are used, for example, for the ion trapping, plasma acceleration, et cetera) but in the 2DEG systems they have been observed, to the best of our knowledge, for the first time in the discussed MIRO/ZRS experiments \cite{Zudov01,Ye01,Mani02,Zudov03}. This has become possible because of the very low scattering of 2D electrons in the GaAs/AlGaAs samples used in those works.

The necessity to work with the ultraclean samples in the MIRO/ZRS experiments hampers the use of these effects in practical applications. We have seen, however, that the value of the ponderomotive forces in the considered systems also depends on the contact properties. In particular, the hard restriction on the mobility of the 2D electrons could be substantially softened in structures with the very thin (in the $z$-direction) contacts, in particular, in graphene systems \cite{Neto09}. The graphene thickness ($\simeq 1$\AA) is about three orders of magnitude smaller than the inhomogeneity scale $l$ that we have assumed above ($l\simeq 0.1$ $\mu$m), therefore the observation of the nonlinear plasma effects, in particular, the microwave induced ponderomotive forces, should be quite possible in the graphene systems. Further studies of such effects may lead to new interesting applications.

So far we have discussed the MIZRS/MIRO effects only in semiconductor GaAs/AlGaAs systems. Recently, a very similar phenomena have been also discovered in the 2DEG systems on the surface of liquid helium \cite{Konstantinov09,Konstantinov10}, in which the electron mobility is also very high. These observations can be also explained by the ponderomotive forces but the interpretation of the effect needs some modifications since the experimental setups and parameters are quite different in the electrons-on-helium systems. In such systems the sample dimensions are large (of the centimeter scale) and the contacts are placed above the 2DEG plane (in fact, the photoresistance in the 2DEG-on-helium systems is measured by a contactless technique, see a typical experimental setup in Fig. 1 in Ref. \cite{Konstantinov08}). Under such conditions the field-amplification effect near the contacts is not the case and the contact-field enhancement factor ${\cal F}_c$ is irrelevant. On the other hand, the 2D electrons are placed inside a cavity, in which the powerful microwave radiation produces a standing wave across the whole area of the system (like in the laser-field induced cold-ion traps in free space). Under the action of the resulting periodic ponderomotive potential the electron density becomes inhomogeneous which leads to the observed microwave induced effects. 

Since in the 2DEG-on-helium systems the factor ${\cal F}_c$ is not big, one could think that the observation of MIZRS effect would require a much higher microwave power as compared to the semiconductor systems, see the MIZRS observability conditions in Eq. (\ref{MIZRSobservability}). This is not true, however, since the electron density in the electrons-on-helium systems is about four orders of magnitude lower than in semiconductors ($\simeq 10^7$ cm$^{-2}$ vs $3\times 10^{11}$ cm$^{-2}$ in GaAs). The formation of the microwave induced electron traps should therefore be easily observable in the 2DEG systems on the surface of liquid helium.

\begin{acknowledgments}

This work would not be possible without numerous discussions, clarifying details of the MIZRS and other microwave experiments, with Ramesh Mani, Jurgen Smet, Klaus von Klitzing, Sergei Dorozhkin and Igor Kukushkin. I would also like to thank Alexei Chepelianskii for interesting comments, Denis Konstantinov for information on the details of experiments \cite{Konstantinov09,Konstantinov10,Konstantinov08} and Nadja Savostianova who has carefully read the manuscript and suggested many useful corrections. 

The financial support of this work by the Deutsche Forschungsgemeinschaft is gratefully acknowledged.

\end{acknowledgments}


\end{document}